\documentclass[9pt]{sigplanconf} 

\pdfoutput=1
\usepackage{graphicx}
\usepackage{amsmath}
\usepackage{amsthm}
\usepackage{amsfonts}
\usepackage{amssymb}
\usepackage{isolatin1}
\usepackage{ulem}
\usepackage{epic}
\usepackage{eepicemu}
\usepackage{boxedminipage}
\usepackage{hyperref}
\usepackage{subfigure}
\usepackage{stmaryrd}
\usepackage[mathscr]{euscript}
\usepackage[all]{xy}  
\usepackage{listings}
\usepackage{xcolor}
\lstnewenvironment{cpp}
    {\lstset{language=c++,fontadjust}}   
    {}

\newcommand{\editor}[1]{{\color{red}\begin{sc}#1\end{sc}}}

\theoremstyle{definition}

\newcommand{\reffig}[1]{Figure~\ref{#1}}
\newcommand{\refsec}[1]{Section~\ref{#1}}

\newcommand{\ignore}[1]{}
\DeclareSymbolFont{AMSb}{U}{msb}{m}{n}
\DeclareMathSymbol{\N}{\mathbin}{AMSb}{"4E}
\DeclareMathSymbol{\Z}{\mathbin}{AMSb}{"5A}
\DeclareMathSymbol{\R}{\mathbin}{AMSb}{"52}
\DeclareMathSymbol{\Q}{\mathbin}{AMSb}{"51}
\DeclareMathSymbol{\I}{\mathbin}{AMSb}{"49}
\DeclareMathSymbol{\C}{\mathbin}{AMSb}{"43}

\newcommand{\paragraphlabel}[1]{}

\newcommand{\tinysection}[1]{\vspace{0.1in}\noindent{\bf #1}}

\newcommand{\future}[1]{}

\newcommand{\iflong}[1]{}

\newcommand{\topicsentence}[1]{#1}

\newcommand{\ifverbose}[1]{}

\begin{document}
%
\conferenceinfo{GPCE'07,} {October 1--3, 2007, Salzburg, Austria.}
\CopyrightYear{2007}
\copyrightdata{978-1-59593-855-8/07/0010}

\title{Parsimony Principles for Software Components and Metalanguages}
\subtitle{[Extended Abstract]}

\authorinfo{Todd L. Veldhuizen}
	{Electrical and Computer Engineering \\ University of Wateroo, Canada}
	{tveldhui@acm.org}

\maketitle
\begin{abstract}
Software is a communication system.  The usual topic of communication
is program behavior, as encoded by programs.  Domain-specific
libraries are codebooks, domain-specific languages are coding
schemes, and so forth.  To turn metaphor into method, we adapt tools
from information theory---the study of efficient communication---to
probe the efficiency with which languages and libraries let us
communicate programs.  In previous work we developed an
information-theoretic analysis of software reuse in problem domains.
This new paper uses information theory to analyze tradeoffs in the
design of components, generators, and metalanguages.  We seek answers
to two questions: (1) How can we judge whether a component is over-
or under-generalized?  Drawing on minimum description length principles,
we propose that the best component yields the most succinct
representation of the use cases.  (2) If we view a programming
language as an assemblage of metalanguages, each providing a
complementary style of abstraction, how can these metalanguages aid
or hinder us in efficiently describing software?  We describe a
complex triangle of interactions between the power of an abstraction
mechanism, the amount of reuse it enables, and the cognitive
difficulty of its use.

\end{abstract}








\section{Introduction}

``{\it Metaphor is an invitation to see the world anew....
Metaphor transfers meaning from one domain into another and thereby enriches
and enhances both domains.}'' \cite{Barrett:JABS:1990}

The design theorist Donald Sch{\"o}n wrote extensively on
the role of metaphor in design.  One of his most famous ideas
is that of {\it generative metaphor} \cite{Schon:MT:1978}, which describes a frame or
perspective carried over from one domain to another to produce
new insights.  A consciously embraced metaphor can be enabling,
but an unacknowledged (tacit) 
metaphor can cast a `spell' over problem solvers, restricting their 
ability to see problems objectively.

Consciously or not, in software engineering we inevitably find ourselves invested in
generative metaphors.  Particularly
prominent is McIlroy's fecund metaphor of ``mass-produced software components,''
which spawned multiple generations of research on software factories,
software product lines, software assembly lines, software robots, and
so forth. 

In this paper we explore the three generative metaphors for
software components:

\begin{enumerate}
\item Software as communication;
\item Component design as statistical model-fitting;
\item Software abstractions as computable functions.
\end{enumerate}

\tinysection{Software as communication.}
\topicsentence{A fruitful viewpoint for understanding the role of 
abstractions in design
is that of software as a communication system.}
The subject of the communication is \emph{program behaviors},
as encoded in a programming language.
Communication systems are a primary object of study in the field
of information theory, and so information theory can rightly be expected to
have much to say about abstractions and their role in design.


\ignore{***
\topicsentence{Information theory studies the efficiency of
communication, and its tools provide a way to study the
efficiency with which programs can represent behaviors.}
***}

\topicsentence{Efficient communication can be achieved by
identifying frequently-occurring patterns or motifs in messages.}  Messages can
then be compressed on average by assigning shorter codes to motifs.
For instance, in spoken English the term ``automobile carriage''
was supplanted in the 20th century by ``car,'' a more efficient
form that reflects its increased use frequency.  These ideas carry
over in a straightforward way to software.
Good software abstractions capture commonly occurring motifs, and we 
can represent our programs more concisely (i.e., compress them) by 
referring to predefined abstractions, rather than describing them anew for each
program.  Abstractions also compress the design process
itself, allowing us to discuss and reason about designs in terms of
recognizable, high-level chunks.

The suggested correspondence between software design and information
theory is summarized by the following table:

\vspace{1em}
\begin{tabular}{ll}
{\bf Software Design} & {\bf Information theory} \\
Problem domain & Random process \\
Program & Message \\
Programming language & Encoding scheme \\
Library & Codebook \\
Abstraction & Motif \\
Code reuse & Compression
\end{tabular}
\vspace{1em}

\topicsentence{In previous work we developed an information-theoretic
view of software libraries \cite{Veldhuizen:LCSD:2005}.}
\topicsentence{This paper extends this work by using an information-theory
perspective to analyze tradeoffs in the design of components and
domain-specific languages.}
We investigate two questions:
\begin{enumerate}
\item In designing a library for a problem domain, how can we evaluate
whether a component is undergeneralized, overgeneralized, or `just right'?
\item How and why should we strike a tradeoff between the power of
abstraction mechanisms in languages, their ease of use, and the amount
by which they allow program length to be reduced?
\end{enumerate}

\ignore{***
\editor{Negotiating tradeoffs.  First, between under- and over-generalization.
Second, between the power of abstraction mechanisms, their ease
of use, and the amount of compression they offer.  Distinguish between
two distinct uses of metaprogramming.  One, to accomplish staging.
Two, as ``abstraction by other means.''}
***}

\ignore{***
\editor{Relationship between natural language grammar, and abstraction
mechanisms?}

\topicsentence{How can we tell if a component is over- or under-generalized?}

\topicsentence{...(the context of this paper wrt the Kolmogorov/Zipf's law
paper)...}

\topicsentence{The MDL principle clarifies the tradeoff in software
design between over- and under-generalization of components.}

\topicsentence{To those who would argue there is no such thing
as too much generalization, I suggest the following counterexample.}

***}

\section{Background}

\ignore{***
This design canon may be materialized as practices, libraries,
conceptual abstractions, notations, etc.  \editor{(Perhaps  go
for a comprehensive list here.)}

\editor{For a design discipline to be effective, (1) there must be
motifs; (2) we must be able to capture those motifs with
abstractions.  (Ah: maybe paste later paragraph `To account for
reuse as we see it in practice, ...')}
***}

\subsection{Information theory of design}

The development of a design discipline around a problem
domain can
be understood as a process whereby people solving design problems
in a new domain identify recurring motifs in their successful
designs, which they abstract into reusable form.
As the domain matures, the most useful abstractions form a
{\em design ca\-non} that represents the core knowledge of the
design community.
From an information-theoretic perspective, the design can\-on is
a \emph{codebook} defined
to \emph{compress} (provide terse representations of) design solutions.

We have proposed that a problem domain be associated with
a probability distribution on programs,
where the distribution reflects the likelihood that someone 
working in the problem domain
will set out to realize a particular program.
Many interesting questions about design can be reduced, via information
theory, to properties of this distribution: the scale of abstraction
at which we can work, the limits of software reuse, and the rates at
which software components can be reused.
For instance, information theory dictates that the extent to which software reuse
can occur is governed by the
`\emph{entropy parameter}' $H$ of this distribution.
In ``low-entropy'' problem domains (with $H$ near $0$),
programs are highly similar to one another and we can design at a high level
of abstraction.
For problem domains with $H$
near $1$, the potential for reuse is low and each program requires substantial
quantities of new code.
We developed this viewpoint in the paper \cite{Veldhuizen:LCSD:2005}.

\tinysection{Tradeoffs in abstraction mechanisms}
Programming languages provide a variety of abstraction mechanisms that
serve to capture common patterns in designs.
Critical to developing good design notations and programming languages
is understanding the tradeoffs between various forms of abstraction
in terms of succinctness, safety properties, complexity, and so forth.
A preliminary investigation of such tradeoffs from the perspective
of computability theory has identified useful avenues of exploration
\cite{Veldhuizen:PEPM:2006}.  
In this work we develop the understanding of such tradeoffs further,
extending our analysis to encompass connections with the psychology 
of programming, 
and develop means to communicate such results to practitioners and language
designers in a meaningful way.
One way to summarize tradeoffs is to sketch ``tradeoff curves'' for 
forms of abstraction,
as commonly used in engineering design.  Such curves give an
intuitive appreciation of the tradeoffs inherent in selecting
forms of abstraction.

\section{Minimum Description Length Principles}

Good component design must strike a balance between over- and
under-generalization.
In this section we describe how
the Minimum Description Length (MDL) principle, pioneered for 
choosing statistical models, also provides a useful framework for 
reasoning about the best level of generalization for a component.

\subsection{The generality problem}

\topicsentence{Generality of software components plays a key role
in their reusability:}
A general-purpose component is more
likely to be reused.  This has been known since the
dawn of time (in software terms): describing in 1952 
the first major subroutine library for the Cambridge
EDSAC, the late David Wheeler wrote:
\begin{quote}
It may be desirable to code [the subroutine] in such a manner that the
operation is generalized ... \cite{Wheeler:ACM:1952}
\end{quote}
\noindent

\noindent
If a component is `not general enough,' we call it
\emph{undergeneralized}.
The development of polymorphism, generics, metaprogramming, 
and so forth has led to languages in which a reasonable
level of generality is easy to achieve.  The ease with which
components can be generalized occasionally leads to the
problem of \emph{overgeneralization}.
Components pay for their generality (or, more correctly,
their users pay) by requiring too much work to
configure, glue, adapt, and so forth.  As a
slightly facetious example we mention that the ultimate 
generalized subroutine in C++ is something such as

\begin{cpp}
template<typename Inputs, 
  typename Outputs,
  typename Operation>
void do(Inputs& in, Outputs& out,
  Operation& op)
{
  op(in,out);
}
\end{cpp}

This function can be made to do most anything one might
want; however, to specialize this function to
a specific purpose, one has to do as much (or more) work
than required by a direct implementation.
(For functional languages, something of the form $\lambda x . \lambda y . xy$
is analogous.)

For a more realistic example, consider matrix multiplication.
The standard Basic Linear Algebra Subroutines (BLAS) function
for this is called DGEMM.  Rather than being merely a 
matrix multiply, DGEMM is generalized to carry out
an operation of the form
\begin{align*}
C &\leftarrow \alpha AB + \beta C
\end{align*}
and optionally, one may transpose any subset of the
matrices.  This broad functionality is paid for by
a somewhat unwieldy interface.  The C binding is:

\begin{verbatim}
void dgemm(char transa, char transb, int m, 
  int  n,  int  k, double  alpha,  double *a, 
  int lda, double *b, int ldb, double beta, 
  double *c, int ldc);
\end{verbatim}

\noindent
The parallel version (from ScaLAPACK) has even more
parameters:
\begin{verbatim}
void pcgemm(char transa, char transb, int m,
  int n, int k, double alpha, double *a,
  int ia, int ja, int *desca, double *b,
  int ib, int jb, int *descb, double beta,
  double *c,  int ic, int jc, int *descc);
\end{verbatim}

Let me emphasize that this is not gratuitous overdesign; there
are situations in which this flexibility is needed.
However, using such interfaces
requires great concentration from the
uninitiate.  Similarly unwieldy interfaces were
commonplace when C++ templates were first
introduced, and the fashion was to anticipate
every possible variation with a template parameter.

Clearly, we must tradeoff
the \emph{generality} of
an abstraction against the difficulty of applying it.  Too specific, and the 
abstraction has limited applicability.  Too general, and it becomes
arduous to adapt.  We call the problem of finding the
right amount of generality for a component `the generality problem.'

We propose a solution for the generality problem based on the
`Minimum Description Length' principle that
has proven so informative in statistics \cite{Rissanen:1989}.
The resulting insights yield practical methods to gauge whether
abstractions are over- or under-generalized.

\begin{figure*}
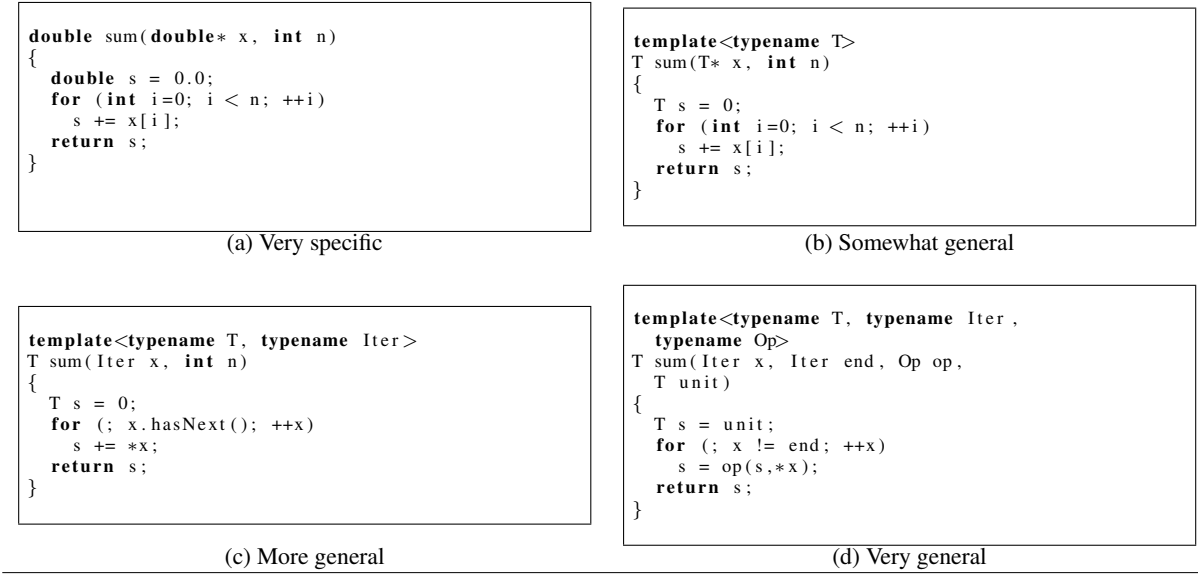

\begin{tabular}{cc}
\begin{boxedminipage}{3in}
\scriptsize
\begin{cpp}
double sum(double* x, int n)
{
  double s = 0.0;
  for (int i=0; i < n; ++i)
    s += x[i];
  return s;
}
\end{cpp}
\vspace{0.17in}

\end{boxedminipage}
&
\begin{boxedminipage}{3in}
\scriptsize
\begin{cpp}
template<typename T>
T sum(T* x, int n)
{
  T s = 0;
  for (int i=0; i < n; ++i)
    s += x[i];
  return s;
}
\end{cpp}
\end{boxedminipage}
\\
(a) Very specific & (b) Somewhat general\\
\\
\begin{boxedminipage}{3in}
\scriptsize
\begin{cpp}
template<typename T, typename Iter>
T sum(Iter x, int n)
{
  T s = 0;
  for (; x.hasNext(); ++x)
    s += *x;
  return s;
}
\end{cpp}
\end{boxedminipage}
&
\begin{boxedminipage}{3in}
\scriptsize 
\begin{cpp}
template<typename T, typename Iter, 
  typename Op>   
T sum(Iter x, Iter end, Op op, 
  T unit)  
{ 
  T s = unit; 
  for (; x != end; ++x)  
    s = op(s,*x);  
  return s;   
}
\end{cpp}
\end{boxedminipage} \\
(c) More general & (d) Very general \\
\end{tabular}
\caption{\label{f:mdl}Four functions that can be used to sum elements of a numeric
array.}
\end{figure*}

As a running example, consider summing the elements of a data 
structure containing numbers.
The C++ function shown in \reffig{f:mdl}a has limited applicability:
it can be applied only to arrays of doubles.
By generalizing this function, we can increase its reuse potential.
In languages
with generics facilities, of which C++ is one, one can `lift' the
function of \reffig{f:mdl}a to a generic algorithm
\cite{Musser:SPE:1994,Bert:ESOP:1988,Goguen:SR:1989,Liskov:CACM:1977}.  As a first step one might abstract over the type of
the array (\reffig{f:mdl}b).
To generalize further, one can abstract over the data structure,
replacing the array with an iterator \cite{Shaw:CACM:1977,Musser:SPE:1994}, as 
in \reffig{f:mdl}c.
One might further generalize over the operation, allowing not just
summation of elements but also multiplication and so forth, shown
in \reffig{f:mdl}d.

\ignore{***
A logical endpoint of this progression, ad absurdum,
would be a subroutine for which one must supply as parameters both 
the data structure and the algorithm to be applied to it.
To apply this version, one submits the body of the desired function 
as the argument $y.apply(\cdots)$, and the data --- in whatever
form it might be--- as $x$.  
Clearly this routine can be applied
in a vast number of scenarios, and yet is not useful since the effort
required to use it is greater than writing a special purpose routine;
in CBSE parlance one might say the size of the glue (adaptation code)
exceeds the size of the component, rendering it useless.
***}

Which of the versions of \reffig{f:mdl} is the right one?  Or,
perhaps, can this question even be asked in a form that is
well-posed and suggests an answer?

\subsection{MDL Principles}

There is a strong resemblance between the generality problem for
software components, and the problem of fitting statistical models
to data.  

A central problem in statistics
is understanding data by fitting a model to it.  Typically
one has a particular \emph{model class} in mind, for example, polynomials.
When the models have many degrees of freedom, the model can fit 
the data too closely--- for instance, finding a high degree polynomial
that passes through every data point exactly.  Such models fail to
capture the underlying character of the data.  

A solid theory of such tradeoffs has been developed by Rissanen
\cite{Rissanen:1989}, and separately by Wallace and his colleagues.
We follow the formulation of Rissanen.
The Minimum Description
Length (MDL) principle states that the
model providing the best explanation
of the data is the one providing the \emph{shortest} explanation of
the data.  
This leads to practical methods for model fitting that balance
the \emph{parsimony} of the explanation against the closeness
of fit to the observed data.  The MDL principle chooses the
model that lets one best compress the data; this implies one
picks the model most adept at finding regularities in the data.


The MDL Principle is as follows:
\begin{quote}
The best explanation of the data is the one that minimizes the sum of
\begin{enumerate}
\item the number of bits required to describe the model; and
\item the number of bits required to encode the data relative to the
model.
\end{enumerate}
\end{quote}

For example, to encode data with a linear regression model, one
would first encode the slope and offset of the line, and then
encode the deviations of the data points from that line.
If this encoding were more succinct than, say, an encoding using
a quadratic model, the linear model would be considered a
better fit.

A similar principle can clarify the tradeoff in software design
between over- and under-generalization.  The proposed correspondence
is as follows:

\vspace{1em}
\begin{tabular}{ll}
{\bf Statistics} & {\bf Software design} \\
Model parameters & Parameters and glue code \\
Model & Abstraction \\
Model class & Abstraction mechanism \\    
Data point & Use scenario \\
Underfit & Under-generalization \\
Overfit & Over-generalization
\end{tabular}
\vspace{1em}

\noindent
We paraphrase the principle for software components:
\begin{quote}
The best level of generality of a software component is that which
minimizes the sum of:
\begin{enumerate}
\item the code length required to implement the component; and
\item the code length required to adapt the component to the desired
use cases.
\end{enumerate}
\end{quote}

\noindent
In the next sections we describe how this can work in practice,
and what the implications of adopting this principle are.

\subsection{Applying the MDL Principle for components}
\label{s:mdlpractice}

The MDL principle requires modification to yield
sensible results for software components.
Our starting point is a set of use cases $U_1,\ldots,U_n$
and a set of candidate components $C_1,\ldots,C_k$.
We assume the abstractions are ordered from least to most
general, in the sense that the functionality of $f_{i+1}$ 
is a superset of $f_i$.

To apply the MDL principle we need an appropriate measure of
code length.  Using bits, as in Rissanen's formulation for
statistics, would select the component that best compressed
the use cases.  However, this would favour cryptically terse
implementations over more readable ones.
What we need is a code length measure that moderates the notion
of succinctness with a nod toward usability.  In preliminary
experiments, we have found that the token count
provides reasonable results.  The token count is invariant
under symbol renamings, comments, whitespace, and so forth,
so that one cannot make a component `better' (with respect
to the MDL principle) by stripping comments and choosing
one-letter variable names.


We have found that the following approach yields
sensible results:
\begin{enumerate}
\item For each component and use case combination, write
code that uses the component to implement the use case.
If the component cannot be adapted to the use case,
then write the simplest possible implementation of the
use case without the component.
\item For each component, count the tokens required to:
\begin{enumerate}
\item implement the component; and
\item adapt it to each use case.
\end{enumerate}
\item The MDL principle, as adapted for components,
suggests that the component minimizing the count of (2)
possesses the `right' level of generality.
\end{enumerate}

\subsection{Example}

We illustrate the application of the
MDL principle by considering three use cases for
the candidate components of \reffig{f:mdl}: 
\begin{enumerate}
\item Summing an array of double-precision floating point numbers
(doubles);
\item Summing an array of integers;
\item Summing an array of floats.
\end{enumerate}

In this scenario many experienced programmers would
opt for the component of \reffig{f:mdl}(b) that
abstracts over the data type.  Abstracting over the
data structure or operation provides no benefit for
these use scenarios, although might be appropriate
depending on anticipated future needs.

To determine what the MDL principle suggests, we
implemented four versions of these use scenarios,
one for each component in \reffig{f:mdl}.  For
\reffig{f:mdl}(a) we merely duplicated the code three
times, and edited to change the datatypes.
For the remaining components we `adapted' them
to each use scenario by providing appropriate
template parameters and arguments.  

\begin{figure}
\scriptsize
\begin{cpp}
/* The component of Figure 2(d) */
template<typename T, typename Iter, typename Op> 
T sum(Iter x, Iter end, Op op, T unit)  
{
  T s = unit; 
  for (; x != end; ++x)  
    s = op(s,*x);  
  return s;   
}

/* Code to adapt to use cases */

/* A function object */
template<typename T>   
struct plus {   
  T operator()(T x, T y) { return x + y; } 
};

/* Three use cases */
double sum_double(double * x, int n)      
{
  return sum(x, x+n, plus<double>(), 0.0);
}

int sum_int(int* x, int n)
{
  return sum(x, x+n, plus<int>(), 0);
}

float sum_float(float* x, int n)
{
  return sum(x, x+n, plus<float>(), 0.0f);
}
\end{cpp}
\caption{\label{f:usecases}The component of \reffig{f:mdl}(d)
with the code required to adapt it to three use cases.
}
\end{figure}

\reffig{f:usecases} shows the code for component
\reffig{f:mdl}(d) and the code needed to adapt it
to the three use scenarios.  Using an automated
tool to count the number of tokens, we obtain the
following results:

\vspace{0.1in}
\begin{tabular}{llll}
 & Component & Adaption & Total \\
Component  & Tokens & Tokens & \\
(a) & 41 & 82 & 123 \\
(b) & 46 & 60 & 106 \\
(c) & 42 & 66 & 108 \\
(d) & 56 & 121 & 177
\end{tabular}
\vspace{0.1in}

\begin{figure}
\begin{centering}
\includegraphics[scale=0.3]{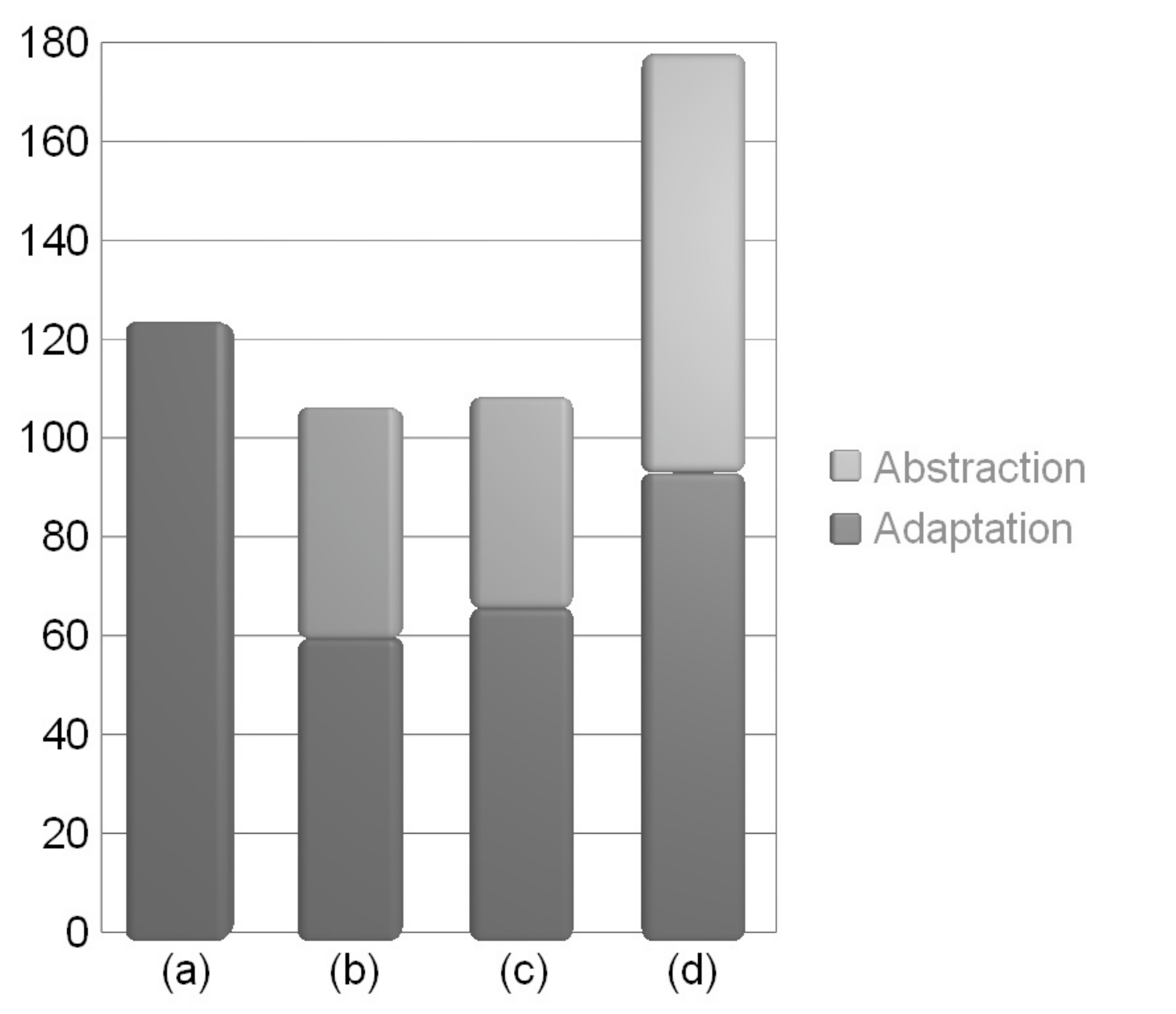}

\end{centering}
\caption{\label{f:mdlgraph} {Applying the MDL principle for generics.  The four bars represent the code size required to use the functions of \reffig{f:mdl} for three use cases.  Code size is measured in tokens.  The light bar indicates the size of the generic function, and the dark bar shows the amount of code required to adapt the generic function to three use scenarios.  In (a), there is no generic function and instead \reffig{f:mdl}(a) is duplicated for each use case, with a different type replacing $\mathsf{double}$.}}
\end{figure}

These results are plotted in \reffig{f:mdlgraph}.  From
the math one expects a convex (U-shaped)
function, and this can be seen in \reffig{f:mdlgraph}.
If the component is too specific, it
cannot be used for all the use cases.
If the component is too general, then
it is arduous to adapt.
And, in between the two extremes, we expect a
component with the `right' level of generality, 
which the MDL principle recommends to us.

\subsection{Discussion}


The proposed MDL principle for components
can be summarized as: `the best component
yields the most succinct representation of the use cases.'

We propose this not as an absolute, but rather as a
guiding principle.  Among the advantages of following
this principle are:
\begin{enumerate}
\item Choosing abstractions according to the MDL principle yields
succinct programs.  If we presume a correlation between
program length and development/maintenance costs, this suggests
following MDL principles in component design would be beneficial.
\item The MDL principle weeds out components that are overgeneralized.
Overgeneralized components typically have numerous parameters
not needed in ordinary usage.  These parameters can make the
component harder to use for novice users, and the possibility
of variation may, in practice, translate into a possibility of
error.
\item The MDL principle provides a quantitative, non-subjective
criterion for choosing the right level of generality.
\end{enumerate}

However, in practice it would be an unreasonable amount of
work to approach the design of every component by applying
the strategy described in \refsec{s:mdlpractice}, i.e.,
implement all the components and use cases and evaluating
the description length.  We do not advocate the MDL principle
as a day-to-day design tool, but rather for the following
uses:
\begin{enumerate}
\item As a teaching tool, a mindset, an exemplar of 
`optimal' design that can be used to guide more informal decision-making.
\item As a retrospective tool with which to evaluate 
existing library APIs, with the aim of extracting general
design principles and recommendations.  In a forthcoming paper we 
apply the MDL principle to evaluating the API of some existing generic
libraries, notably the STL, with use cases gleaned from
open source projects.  Our preliminary results
suggest the STL is overgeneralized with respect
to the use cases for which it is most commonly employed.
\end{enumerate}

Finally, we note that the MDL principle suggests that the
`right' level of generality can only be gauged with respect
to a set of use cases.  These use cases might be chosen
to guide the design of a component for a specific project,
or they may be chosen to represent typical usage in a problem
domain.  In the latter case we can think of the use cases
as sampling the distribution of programs associated with
the problem domain.

\section{Abstraction mechanisms}

\ignore{***
\editor{
The abstraction mechanisms provided by programming
languages determine our ability to represent programs succinctly.
When we exhaust the capacity of a language to abstract, we turn
to metaprogramming.
Metaprogramming is simply a continuation of abstraction, with the
addition of other means.
}
***}




\topicsentence{Modern programming languages support multiple forms of 
abstraction,
each with a distinct sublanguage in which they are defined.}  For
example, in the language C++ there are distinct sublanguages for
defining classes, functions, generic functions, and macros.  The
evolution of programming languages can be seen in part as an ongoing
quest to identify useful forms of abstraction and formalize them
as language features.
\ignore{***
Some forms of abstraction are not codified by languages, but are formalized
as design practices, for example, design patterns \cite{Gamma:SP:2002}.  
Others persist as informal tricks and techniques in the community.
***}

\ignore{***
Viewing software as communication leads to the beginnings of a theory of
how to best apply different forms of abstraction to manage complexity
in design, and what kinds of abstraction deserve or require formal
support.
}

The difficulty of \emph{adapting}
or \emph{instantiating} an abstraction for a particular use scenario
appears to relate directly to the computational complexity of the
problem of \emph{inverting} the abstraction.  The cognitive difficulty
of spotting common motifs in programs that can be abstracted away
appears to be closely related to the computational complexity of
\emph{compressing} the program with respect to a class of abstractions.

In this section we explore the tradeoff between the power of
abstraction mechanisms, the degree of succinctness they offer,
and the cognitive difficulty of their use.

\subsection{Metalanguages as classes of functions}

\topicsentence{
To examine the differences between forms of abstraction, we need a common 
framework in which to compare them.}  We believe a useful viewpoint is that
of abstraction mechanisms as classes of functions, in particular, as
classes of partial computable functions.
The rationale is as follows.  An abstraction represents
a set of concrete \emph{instances}.  For example, a generic linked list class
$\mathsf{List}\langle T \rangle$ can be \emph{instantiated} to 
instances such as $\mathsf{List}\langle \mathsf{int} \rangle$ and
$\mathsf{List}\langle \mathsf{string} \rangle$; we can associate with
$\mathsf{List}\langle T \rangle$ a function mapping the parameter $T$
to instances.  Similarly, 
a macro can be associated with a map that substitutes parameters into
  the macro definition;
a subroutine can be associated with an abstraction function that 
  substitutes arguments for variables and inlines the function body;
a class definition can be associated with a function
  that imbues subclasses with its functionality;
a parser generator can be associated with a map from grammar specifications
  to parser implementations.

This gives abstraction mechanisms an operational interpretation, 
e.g., the activity a compiler would
carry out to reduce the abstraction to a lower-level representation.
Alternately, we can think of abstraction mechanisms as defined by
a denotational meaning, e.g., we associate with each abstraction 
a function from parameters to object language semantics.
In either case, we can associate with each abstraction
some \emph{abstraction function} that gives it meaning,
either operationally or denotationally.

\ignore{***
\editor{For thinking about inversion problems + cognitive difficulty,
probably want to think of an abstraction function as being from terms
to denotational semantics of the problem domain.  E.g., the function
hypot would really go to ``$\sqrt{x^2+y^2}$'' or ``Euclidean distance.''}
***}

An abstraction \emph{mechanism} can then be viewed as a class of abstraction
functions, as enumerated by some restricted language we call 
a \emph{metalanguage}, following the usual terminology of metaprogramming.
Programming languages can be regarded as an assemblage of metalanguages,
each offering a distinct form of abstraction.\footnote{For historical
accuracy one might regard a programming language as a pastiche of
metalanguages.}

To fit metalanguages into the framework of computability, complexity
theory, and subrecursive languages, we use the following
correspondences:

\vspace{1em}
\noindent
\begin{tabular}{ll}
{\bf Software design} & {\bf Theory idea} \\
Abstraction & Partial computable \\
& function \\
Abstraction mechanism & Class of p.c. functions, \\
                      & a \emph{metalanguage} \\
Parameters/glue code & Input \\
Instantiation & Evaluation of p.c. function\\
Instance of an abstraction & Output of p.c. function\\
Adaptation & Inversion of p.c. function
\end{tabular}
\vspace{1em}

\subsection{Facets of abstraction mechanisms}

The fact that programming languages provide a variety of abstraction
mechanisms (i.e., metalanguages) suggests that there is no single best
`universal abstraction mechanism.'  Instead we find that
metalanguages offer a broad variety of tradeoffs between
desirable facets, namely:
\begin{itemize}
\item The expressive power of the metalanguage, i.e.,
what abstractions are definable in it.
\item The safety properties we are guaranteed about instances.  For
example, an ongoing
concern in programming language design is finding metalanguages that can
generalize over types in a safe way, e.g., generics \cite{Garcia:OOPSLA:2003}.
\item Succinctness, that is, how long the description of an abstraction
must be, and how long parameters must be to produce instances of interest.
\item The time and space complexity of instantiating an abstraction
(i.e., how intensive the compilation process must be.)
\item The difficulty of finding parameters to an
abstraction that will produce a particular instance, i.e.,
\emph{inversion} of an abstraction.
\item The effort required to devise an appropriate
abstraction, given an instance or class of instances over which 
we wish to generalize.
\end{itemize}

\noindent
In previous work we used tools from computability theory
and the theory of subrecursive languages
to study tradeoffs between succinctness (code length) and
safety properties \cite{Veldhuizen:PEPM:2006}.  
\ignore{***
One of the key ideas is to
understand a metalanguage as a (computably enumerable) set of programs
in a fixed, universal language.  This is made possible by a ``two-part
code'' construction (interpreter plus program), as used in Kolmogorov
complexity \cite[\S 2.1.1]{Li:1997}.  We can then view metalanguages as
restricted subsets of the universal language, which substantially
simplifies the formalization.
***}

In the present work we examine tradeoffs between the 
expressive power of abstractions, the amount of `compression'
they allow, and the cognitive effort required to use them.

\subsection{Tradeoffs and cognitive tasks of design}

In designing a compression algorithm, one is
is interested in the tradeoff between the degree of compression
achieved and the computational cost of compression and decompression.  
In programming languages,
humans do the compression and compilers do the decompression,
so to speak.
In designing a programming language, the tradeoff is largely between
the succinctness a language offers (i.e., amount of compression)
and the cognitive difficulty of recognizing and exploiting motifs
(i.e., the cost of compression).


Perhaps not surprisingly, the difficulty of cognitive tasks we
encounter in design appears to correlate with the computational
complexity of abstraction mechanisms \cite{Wilson:unpub:2005}.
For instance, a crucial design activity
is deciding whether an abstraction can be adapted to
a use scenario.  To support rapid design work, the
cognitive task ought to be simple --- abstractions that require great
deviousness to adapt are unlikely to be used frequently.
The equivalent problem in our formalism
is deciding whether there exist parameters for the abstraction function
that will cause it to produce a desired output --- the problem of
\emph{inverting} the abstraction function (not to be confused with
inverting a runtime computation, an altogether different problem.)
\begin{figure}
\begin{minipage}{2.9in}
\begin{cpp}
double hypot(double a, double b)
{
  return a*a + b*b;
}
\end{cpp}
\end{minipage}
\caption{\label{f:hypot} A simple abstraction that is easy to ``invert.''}
\end{figure}

For example, macros and subroutines are forms of abstraction
``instantiated'' by substitution of arguments for variables.
Consider the function $\mathsf{hypot}$ of
\reffig{f:hypot}.
For the function $\mathsf{hypot}$ to be useful, it must be possible for us
to recognize places in our design where it might be used, and to determine what
parameters will make it do what we want: given the fragment
\begin{align*}
\mathsf{double}~d &= c + r*r + f(s)*f(s);
\end{align*}
it is easy to see that the term $r*r+f(s)*f(s)$ can be replaced by
$\mathsf{hypot}(r,f(s))$.
This can be understood as inverting the substitution process.
The inverse of substitution
is unification, which can be performed in almost linear time \cite{Knight:CSUR:1989}.
It seems significant that \emph{almost all the abstraction mechanisms
we find useful in practice lie in low computational complexity classes, and
are easy to invert.}
This suggests that we tend to favor simple, easy-to-reuse design abstractions
over more complicated (but possibly more general) ones.
Thus, for example, method invocations and inheritance, the workhorses of
object-oriented programming, both appear easy to invert.
On the other hand, arbitrary program generation (e.g., as in
staged languages \cite{Sheard:unpub:1994,Taha:TCS:2000}, generative programming \cite{Czarnecki:2000}, and
template metaprogramming \cite{Veldhuizen:CRb:1995,Alexandrescu:2001,Abrahams:2004})
tends to be used sparingly and often in only simple ways.


\ignore{***
\editor{
Perhaps need to identify the space in which we are reasoning.  With
many abstractions we need to reason on `both sides of the abstraction.'
(i.e., thinking about parameters and thinking about the instantiation.)
With parser generators and compilers, we very rarely need to reason
about the code that is being generated.}
***}

\subsection{Viscosity and Lipschitz abstractions}

Cognitive tradeoffs in design notations
have been summarized by Thomas Green and colleagues in
the popular \emph{Cognitive Dimensions of Notations} framework
\cite{Green:HCI:1989,Blackwell:CT:2001}.  Green argues that
programming languages are properly regarded as a med\-ium in which we hash
out design decisions, not just record them after the fact.  Human design
work --- even that of experts --- has been shown in
numerous studies to be disorderly,
characterized by false starts, frequent rewriting,
and simultaneous attacks on the problem at many levels of abstraction:
``design is redesign, programming is
reprogramming.'' \cite{Green:HCI:1989}.  To support the way
humans design, notations must be malleable---
it must be possible to quickly evolve code to match our
changing understanding of the design.  In the cognitive dimensions
framework this quality is dubbed \emph{viscosity}:
the resistance of a notation to change.
One way to evaluate the `viscosity' of an abstraction mechanism is to
analyze how sensitive the input of the abstraction function is to
small changes in its desired output.
Returning to our earlier $\mathsf{hypot}()$ example, consider a
small change in the use scenario from $r*r + f(s)*f(s)$ to
$r*r + f(s+1)*f(s+1)$.
This change that requires only a minor change to the
parameters: from $\mathsf{hypot}(r,f(s))$ to
$\mathsf{hypot}(r,f(s+1))$.  

\begin{figure*}

\begin{align*}
\xymatrix {
\mathsf{Abstraction ~~Parameters} & & \mathsf{Instance} \\
x=(r,f(s)) \ar[rr]^{\mathsf{hypot}} \ar[d]_{d(x,x')} & & y=r*r+f(s)*f(s) \ar[d]^{d(y,y')}\\
x'=(r,f(s+1)) \ar[rr]^{\mathsf{hypot}} & & y'=r*r + f(s+1)*f(s+1)
}
\end{align*}

\caption{\label{f:distance}{Illustration of how abstractions interact with
edit distance, ideally: a small change in the instantiation (from $y$ to $y'$)
can be achieved by a small change in the parameters (from $x$ to $x'$),
i.e., $d(x,x') \leq d(y,y')$.}}
\end{figure*}
This can be formalized by examining the relation between tree edit distance
\cite{Klein:ESA:1998,Zhang:SIAMJC:1989} of the inputs and
outputs to the abstraction function.  Roughly speaking, tree edit distance
measures how many ``editing operations'' would be required to transform a
term $t$ to a term $t'$, giving a distance metric $d(t,t')$ on terms.
If making small changes in the instantiated code requires large changes
to the parameters,
we expect the notation to be ``viscous'' in the sense of Green, resisting
our efforts to evolve our design.  The cognitive dimensions framework
suggests that
\emph{small changes in the instantiation should be realizable by
small changes in parameters.}  This is illustrated in \reffig{f:distance}.

We can formalize this intuition in terms of
Lipschitz continuity.  A Lipschitz condition on a real function 
$f : \R \rightarrow \R$ is a requirement of the form
\begin{align*}
|f(a) - f(b)| &\leq K|a-b|
\end{align*}
where $K \geq 0$.
A function satisfying this condition is said to be \emph{Lipschitz},
and $K$ its \emph{Lipschitz constant}.  The notion generalizes easily
to metric spaces: given a metric space $(T,d)$, and a function
$f : T \rightarrow T$, we can call $f$ Lipschitz if
\begin{align*}
d(f(a),f(b)) &\leq K d(a,b)
\end{align*}
for some $K \geq 0$.  Viewing abstractions as functions from terms to
terms, and tree edit distance as the distance metric, we can make
the following conjecture:

\begin{quote}
The ease with which an abstraction can be used in design work
is strongly influenced by whether it is Lipschitz, and if so,
the magnitude of its Lipschitz constant.
\end{quote}

Again, it seems significant that the abstraction mechanisms we find useful
in practice usually satisfy this requirement.  For instance, with
substitution (the abstraction mechanism for subroutines), the edit distance
between parameters is at most the edit distance between
instantiations.
This property does not hold in general for
abstraction mechanisms that lie in higher computational complexity classes, so again we return to the observation that useful abstraction mechanisms tend to
be computationally very simple.


\ignore{***
\editor{
Idea of random process model vs. degree of compression that can
be achieved.  Turn this around and look at how compressible a
signal is wrt the amount of effort one is willing to put in.
(Relativized notions of randomness, e.g. Martin-L{\"o}f, resource-bounded
Kolmogorov complexity.)
}
***}

\subsection{Tradeoff curves}
The formalization of abstraction mechanisms as metalanguages 
lends itself to understanding tradeoffs
between forms of abstraction: between their generality and ease-of-use,
between safety properties and succinctness, for example.  In
engineering it is common to sketch tradeoff curves between opposing
factors to aid in choosing the `sweet spot' where a suitable balance
is reached.  We can draw such curves for abstractions also, and we
believe these curves provide an intuitive tool for understanding
tradeoffs.

The tradeoff we are interested in here is that between
\begin{enumerate}
\item the complexity of the abstractions used;
\item the degree to which programs can be `compressed' using those
abstractions;
\item the cognitive difficulty of using ($\approx$ inverting) abstractions.
\end{enumerate}

Intuitively, if we put `abstraction complexity' on an $x$-axis,
we expect that the program length we can obtain decreases
as we increase the complexity of our abstractions, and the
cognitive difficulty increases.

To draw such tradeoff curves in a meaningful way,
we need a few justifications; it is not immediately clear,
for example, how one can put `abstraction complexity' on the
$x$-axis in a meaningful way.
We need a
suitable mapping from metalanguages (e.g., classes
of partial computable functions) to points on an $x$-axis.  
If we have a chain of metalanguages of increasing complexity,
we can map this chain onto an axis by appealing to the classical
result of Cantor that every dense total order without
endpoints is order-isomorphic to $\Q$, the rational numbers.  
This suggests we can take a set of metalanguages, excise a substructure that
is a total order (possibly dense), and embed it in the rationals.
This provides a clean, if somewhat roundabout, explanation for
drawing an $x$-axis of metalanguages.
This correspondence is not unique, and so the placement of particular
elements on the axes is arbitrary (up to ordering); the observed
shape of the graph is meaningful only up to squeezing and stretching
of the axes.  Indeed even the shape of the curve itself is usually
a sketch based on scanty information, and is intended to convey
intuition rather than exact information.

To formalize the notion of `achievable program length,' we could
appeal to either Yao-pseudoentropy \cite{Barak:LNCS:2003}
or a nonuniform variant of resource-bounded Kolmogorov complexity.

To formalize `cognitive difficulty,' we assume a correspondence
between the time complexity of the inversion problem and its
cognitive difficulty.

\begin{figure}
\begin{centering}
\includegraphics[width=\columnwidth]{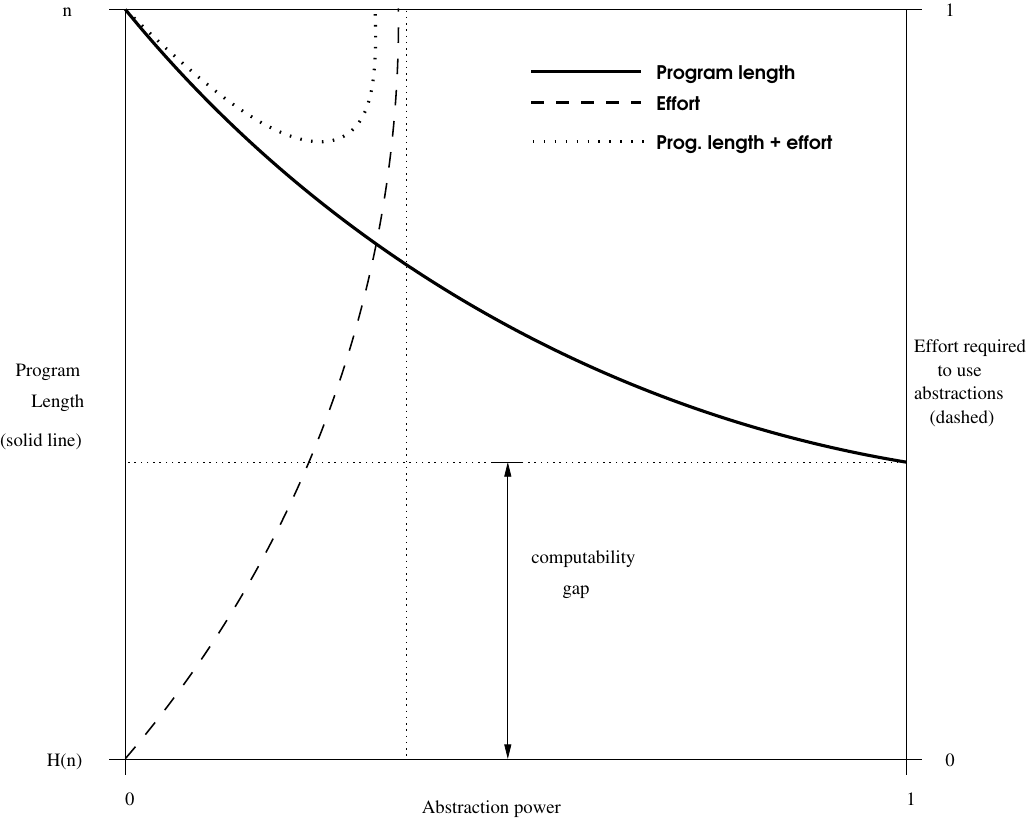}

\end{centering}
\caption{\label{f:compress}\footnotesize Tradeoffs between abstraction power,
component reuse, and difficulty of reusing components.
\vspace{0.15in}
}
\end{figure}

This is all a little vague, and perhaps the analogy is strained.
But such graphs can tell useful stories.
\reffig{f:compress} illustrates the tradeoff between
the complexity of abstractions, their ability to reduce program length,
and their ease of use.
The horizontal axis represents \emph{abstraction complexity},
here normalized so that $0$ represents no power whatsoever, and
$1$ represents unrestricted power, e.g., Turing-complete
program generators.
The left vertical axis shows the expected program size achievable.
As abstraction power increases, so does the scope for component reuse,
and the expected program size
(thick line, left axis) decreases, tending to some optimal value greater
than $H(n)$, the entropy for the problem domain \cite{Veldhuizen:LCSD:2005}).
It is possible we cannot achieve the maximum possible compression $H(n)$ 
because there might be patterns in programs which are not exploitable
in any effective way, leading to what is labeled the ``computability gap''
in \reffig{f:compress}.  
To achieve the best possible compression, this curve suggests we ought
to use a high level of abstraction power, i.e., arbitrary program
generators.
On the other hand, as the power of
abstractions increases, the difficulty of using them in
a given situation
(the complexity of the \emph{inverse problem}) increases
rapidly, quickly becoming noncomputable (dashed line, right axis). Thus we have
a tradeoff between the power of abstractions to generalize, and the
difficulty of adapting them to a particular use scenario.  
The dotted line shows a tradeoff curve with a hypothetical
`sweet spot' that balances the complexity of abstractions
against the program length achievable.

This graph illustrates
why in practice we tend to use computationally weak forms
of abstraction, and use complex forms of abstraction (e.g., program
generators) sparingly, even though they might \emph{in principle}
allow us to achieve much shorter program lengths.



\nocite{Carette:ISSAC:2004}

\section{Conclusions}

We have proposed using the MDL principle to
answer the `generality problem,' of how one
chooses the right level of generality for a software component.
As applied to software components, the MDL principle suggests
that `the best component yields the most succinct representation
of the use cases.'  In forthcoming work we use this approach to
retrospectively evaluate the interface design of generic
libraries.

The second portion of this paper suggested an approach to
understanding the tradeoff in programming language design
between the power of abstraction mechanisms, their ability
to reduce program length, and the cognitive difficulty
of their use.  We observed that almost all the abstraction
mechanisms popular in practice lie in low computational
complexity classes.  A plausible explanation for this
is that such mechanisms are easy to `invert,' e.g.,
we can readily figure out what parameters to provide a
macro to achieve a desired result.  We connected Thomas
Green's notion of \emph{notational viscosity} to the
theoretical notion of Lipschitz continuity, which
formalizes the bridge between cognitive difficulty
and computational difficulty.  Finally, we summarized
the tradeoffs in abstraction mechanism design by
sketching a curve illustrating the `sweet spot' 
that balances the complexity of abstractions,
their cognitive difficulty, and the amount by which
they reduce program length.

\acks

The question of how to decide when components are over- or
under-generalized was posed to me by Sibylle Schupp.

\bibliographystyle{abbrv}

\end{document}